\documentclass[epjST]{svjour}
\usepackage{graphics}

\usepackage{amssymb} 
\usepackage{amsmath} 
\usepackage{subfigure}
\usepackage{epsfig}
\usepackage{graphicx}

\begin{document}
%

\title{Free Cooling Phase-Diagram of Hard-Spheres 
with Short- and Long-Range Interactions}
\author{S. Gonzalez\inst{1}\thanks{\email{tsuresuregusa@gmail.com}} 
\and A. R. Thornton\inst{1}\inst{2} 
\and S. Luding\inst{1}\thanks{\email{s.luding@utwente.nl}}}
\institute{Multiscale Mechanics (MSM), CTW, MESA+, University of Twente, 
P.O. Box 217, 7500 AE Enschede, The Netherlands
\and
           Mathematics of Computational Science, Dept. of
  Appl. Math., University of Twente, P.O. Box 217, 7500 AE Enschede,
  The Netherlands}

%
\abstract{We study the stability, the clustering and the phase-diagram
  of free cooling granular gases. The systems consist of mono-disperse 
  particles with additional non-contact (long-range) interactions, and 
  are simulated here by the event-driven molecular dynamics algorithm 
  with discrete (short-range shoulders or wells) potentials 
  (in both 2D and 3D). Astonishingly good agreement is found with a 
  mean field theory, where only the energy dissipation term is modified to
  account for both repulsive or attractive non-contact interactions.
  Attractive potentials
  enhance cooling and structure formation (clustering), whereas
  repulsive potentials reduce it, as intuition suggests. 
The system evolution is controlled by a single parameter: 
the non-contact potential strength scaled by the
fluctuation kinetic energy (granular temperature).
When this is small, as expected,
the classical homogeneous cooling state is found. However, if the effective
dissipation is strong enough, structure formation proceeds, before
(in the repulsive case) non-contact forces get strong enough to undo the 
clustering (due to the ongoing dissipation of granular temperature).
  For both repulsive and attractive potentials, in the homogeneous regime, 
  the cooling shows a universal behaviour when the (inverse) control parameter 
  is used as evolution variable instead of time.  
   The transition to a non-homogeneous regime, as predicted by stability 
   analysis, is affected by both dissipation and potential strength.
   This can be cast into a phase diagram where the system changes
   with time, which leaves open many challenges for future research. 
} 
\maketitle

\section{Introduction}

Granular gases are granular materials where the duration of a
collision is much shorter than the typical collision time
\cite{haff83,goldhirsch93,mcnamara93b,mcnamara96,herrmann98,luding99,poschel01,brey08}.
This situation can be obtained by either placing a dilute particle
system in a micro-gravitational environment (e.g.\ during a parabolic
flight \cite{Bannerman2011}), or experimentally easier, by feeding the
system with energy such that a gaseous steady state appears (e.g.\ by
vertically vibrating the enclosure \cite{olafsen99,luding03c}). For a granular
gas, in the dilute limit, binary collisions dominate over multiple
collisions. Contrary to molecular gases, granular gases are
dissipative. So the continuous loss of kinetic energy due to collisions
not only makes the gas cool down but is also accompanied by collective
phenomena such as cluster formation or shear banding. Granular gases
are subject to instabilities and cluster formation 
\cite{mcnamara93,mcnamara93b,olafsen99,luding99,miller04t,miller04b,luding05e}, 
deviations from the Maxwell-Boltzmann velocity distribution
\cite{brilliantov00d,brilliantov03}, phase transitions \cite{esipov97}
and the formation of vortices \cite{mcnamara93b}.

The instability that leads to cluster formation is
exclusively an effect of dissipation during collisions for the case of
hard spheres \cite{goldhirsch93}, and thus should be enhanced by
attractive potentials and diminished by repulsive potentials between
the particles, examples of which are electrically charged or self-
gravitating particles \cite{muller08t,muller11}. Such particle systems with
attractive long-range interactions can be found, for example, in dry
powders, electrostatic coating processes or in space. In the latter
case, huge mass distributions of interstellar dust clouds, or dense
granular rings and disks around central bodies can be affected by
considerable self-gravitation \cite{Brahic1976,bridges84,Horanyi1993,poschel01}.
Electrically charged granular media are, in nature and
industrial processes, the rule rather than the exception (see
Ref.\ \cite{Pahtz2010} and references therein). 
However, in real systems, purely repulsive potentials are rare; 
in general there is an attractive well inside the repulsive barrier, which
can lead to stable clustering of a granular gas \cite{dammer04}.

This study is devoted to the behaviour of free cooling granular media 
as studied, e.g.\ by Luding and Herrmann \cite{luding99}, 
with additional non-contact repulsive or attractive potentials,
as reported also in the PhD theses by 
S. Miller \cite{miller04t} and M.-K. M\"uller \cite{muller08t}. 
Here we focus on the dilute limit and the interactions are identical 
for all particles, (i.e.\ attractive 
and repulsive potentials are not active at the same time
and neither are di-polar, e.g.\ magnetic, interactions 
considered \cite{blair03,blair04}). 
Simulations are compared to the mean-field theory by M\"uller
and Luding \cite{muller11,muller09}, which only features a
modified cooling due to either repulsion or attraction.
Our goal is to understand how dissipation and non-contact
interactions act together, and what kind of dynamics is
caused by the interplay of both mechanisms active at the 
same time in a free cooling granular system.


In section\ \ref{sec:granularGas} we present the 2D hydrodynamic
equations for our system. We introduce the modified cooling rate due
to long-range interaction in Sec.\ \ref{sec:coolingRate}. We introduce
our simulation method in Sec.\ \ref{sec:simulations}, and continue
with numerical results in Sec.\ \ref{sec:results}. A preliminary 
phase-diagram is introduced in Sec.\ \ref{sec:phaseDiagram} and the conclusions 
and perspectives of future work finalise the paper in Sec. \ref{sec:conclusions}.

\section{Classical Granular Gas}
\label{sec:granularGas}

In this section we review the theory of the free cooling granular gas,
which is a nice reference case 
since it is analytically solvable under a few simplifying assumptions,
as e.g.\ homogeneity that leads to the homogeneous cooling state
(HCS), and is known in the literature as Haff's law \cite{haff83}. 

\subsection{Homogeneous Free Cooling Theory}
A free cooling granular gas in its HCS \cite{haff83} dissipates
kinetic energy $K$, as governed by the equation
\begin{eqnarray}
\label{eq:haffLaw2}
K(\tau)/K(0) = (1 + \tau)^{-2}~,
\end{eqnarray}
with the rescaled time 
$\tau = (1-r^2)t / [2\mathfrak{D}t^0_E]$. 
$\mathfrak{D}$ is the dimension of the system,
$r$ the coefficient of restitution, and $t^0_E$ 
the initial Enskog collision rate \cite{luding05e,luding09},
\begin{eqnarray}
t_E = \frac{d\sqrt{\pi}}{2^{\mathfrak{D}} \mathfrak{D} \nu g_{\mathfrak{D}}(\nu)\sqrt{T_g/m}}~,
\end{eqnarray}
with the (time-dependent) granular temperature $T_g = \frac{2}{
  \mathfrak{D}}\frac{K(t)}{N}$, i.e., twice the kinetic energy per
particle per degree of freedom, while $T = (\mathfrak{D}/2)T_g/m$ 
is the square of the velocity fluctuations. 
The diameter of the particle is $d$, $\nu$ is
the packing fraction of the system and $m$ the mass of the
particles. 
In 2D, the pair correlation
function at contact $g_{\mathfrak{D}}$ is given, approximately for low
to moderate densities, by \cite{carnahan69,henderson75} \[g_2(\nu) =
\frac{1-7\nu/16}{(1-\nu)^2}~,\] and in 3D by \[g_3(\nu) =
\frac{1-\nu/2}{(1-\nu)^3}~.\] Improved formulae for higher density can
be found in \cite{luding09,torquato95,ogarko2012,gonzalez2010}.

\subsection{Hydrodynamic Equations}
The hydrodynamic equations for a granular gas are explained in detail
in \cite{gonzalez2010}. For the sake of brevity, we present a
summarised version.

The continuity equation for the mass density, ${\rho}$, reads:
\begin{eqnarray}
\label{eq:PDEmass2}
  \frac{D {\rho}}{D {t}} 
  + {\rho} \frac{\partial {u}_i}{\partial {x}_i}
  = 0~,
\end{eqnarray}
where $u_i$ is the velocity in the $x_i$ direction. Momentum
conservation in absence of gravity gives:
\begin{eqnarray}
\label{eq:PDEmomentum2}
  {\rho} \frac{D {u}_i}{D {t}}
  = - \frac{\partial {\sigma}_{ij}}{\partial {x}_j}~,
\end{eqnarray}
where $\sigma_{ij}$ is the stress tensor. The energy balance reads:
\begin{eqnarray}
    {\rho}\frac{D}{D{t}} {T} = - {\sigma}_{ik}
    \frac{\partial {u}_i}{\partial {x}_k} - \frac{\partial
      {q}_k}{\partial {x}_k} - I~,
\label{eq:PDEenergy2}
\end{eqnarray}
where $q_k$ is the heat flux and $-I=-I_0=-\hat{\gamma}T$ the energy 
density dissipation rate in the absence of any additional forces, i.e.\ with
potential energy $\phi=0$. 
For the explicit formulation of the coefficients see
\cite{miller04t,luding05e,luding09,gonzalez2010}.

The HCS for a freely cooling gas is found by taking all the spatial
derivatives in the hydrodynamic equations equal to zero, leaving just
one equation for the temperature. The remaining fields -- density and
velocity in each component -- remaining homogeneous and constant: 
$\nu = \nu_0, ~ \vec{u} = 0$\,. Knowing that $\hat{\gamma}\propto
\sqrt{T}$, and rescaling time, the result is directly
Eq.\ (\ref{eq:haffLaw2}). This solution to the simplified
system of equations is what Haff derived from simple mechanical 
arguments in his seminal paper \cite{haff83}.

\subsection{Cluster Instability}
The homogeneous cooling state, however, is not always a stable
solution for the system. For the hard-sphere potential, when the
system size is large enough (at a given dissipation), the homogeneous
cooling becomes unstable and shear and clustering modes appear in the
system. For the attractive potential we expect that the cluster
instability will be always enhanced, and on the contrary, for the
repulsive cased reduced.

The spontaneous formation of clusters in a force-free cooling granular
gas can be understood by simple arguments
\cite{goldhirsch93,poschel05b}: consider density fluctuations in an
otherwise homogeneous granular gas. In denser regions the particles
collide more frequently than in more dilute regions, therefore, dense
regions cool faster than dilute regions and thus the local pressure
decays faster as well. The resulting pressure gradient causes a flux
of particles into these regions of higher density, 
which leads to further increase of
the density. Hence, small fluctuations of the density are enhanced,
which leads to the formation of clusters.

\begin{figure}[htp]
  \epsfig{file=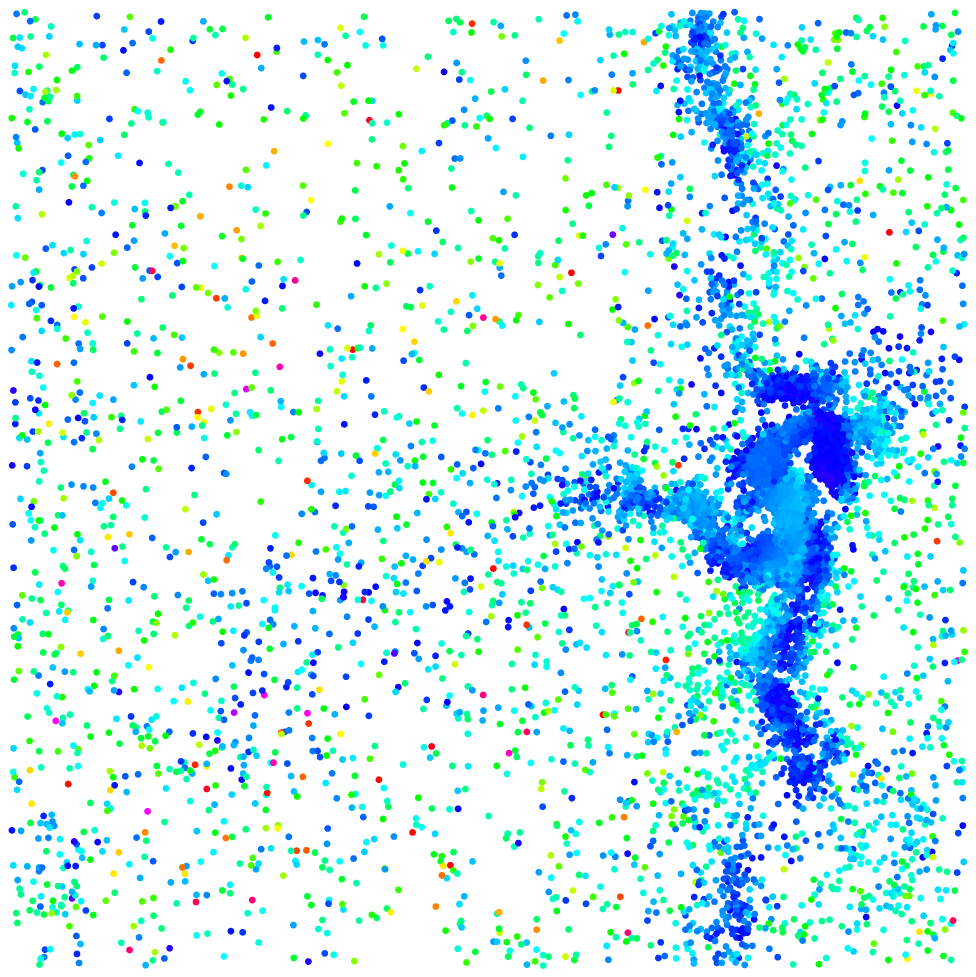,width=.3\columnwidth}
  \hfill
  \epsfig{file=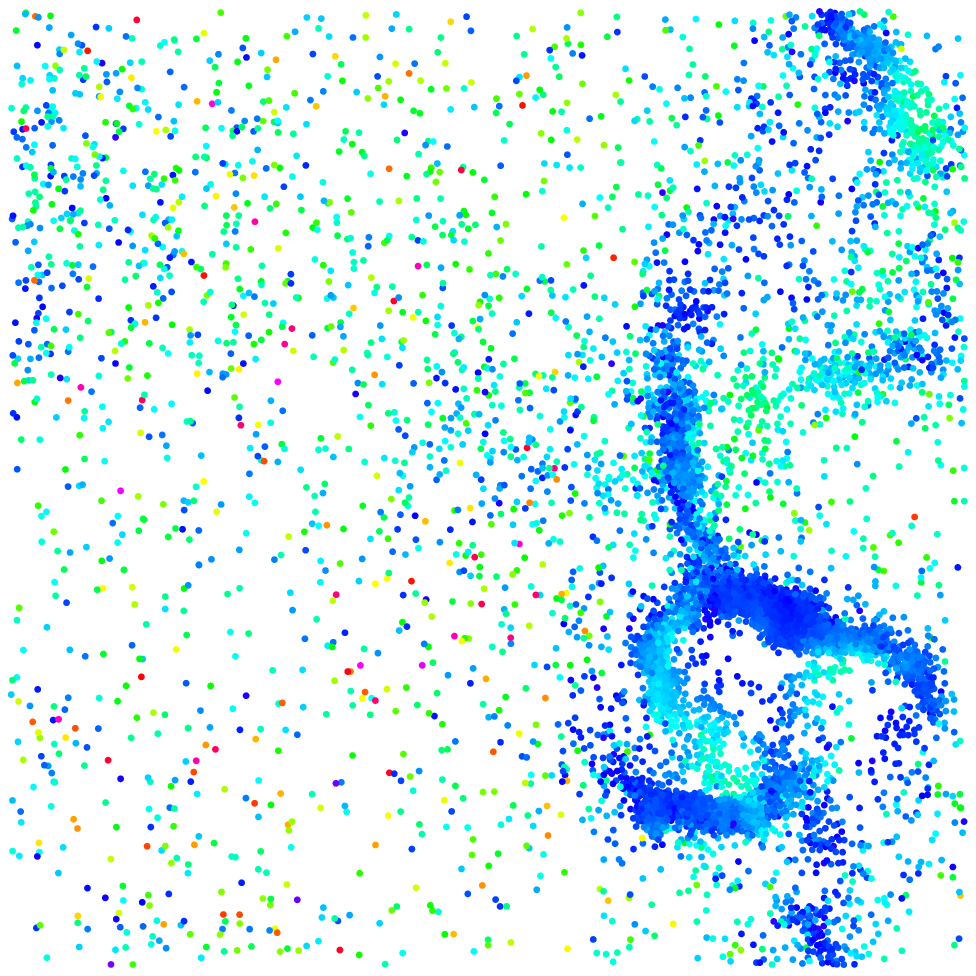,width=.3\columnwidth}
  \hfill
  \epsfig{file=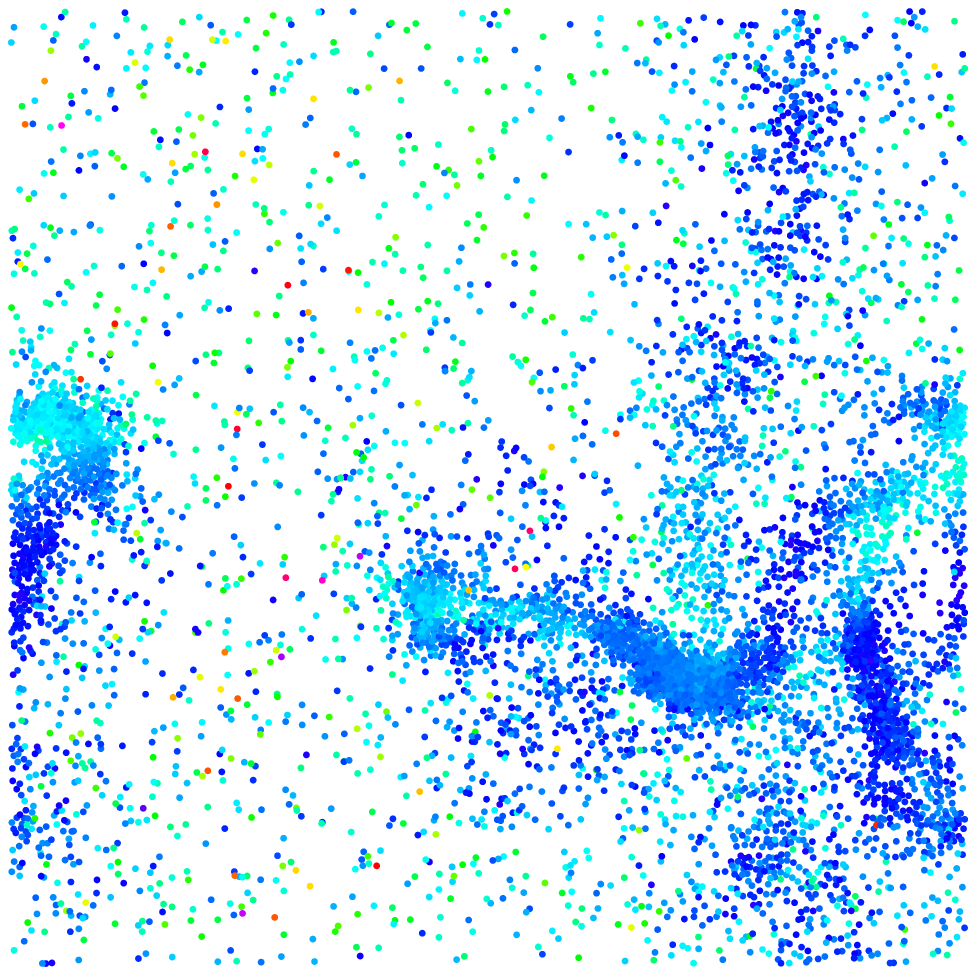,width=.3\columnwidth}

  \caption{Snapshots after $7\times 10^5$ collisions for systems with
    weakly attractive, neutral, and weakly repulsive potentials, from
    left to right. Number of particles $N=6400$, density $\nu=0.0578$
    and dissipation $r=0.65$. The colour code indicates the kinetic
    energy per particle, from blue (slow), to yellow (average) and red
    (fast).}
  \label{fig:3clusters}
\end{figure}

Simulation and hydrodynamic equations are both non-dimensionalised,
with the particle radius and the initial granular temperature. The
potential strength is scaled by the initial too temperature and hence
is dimensionless. Fig.\ \ref{fig:3clusters} shows snapshots for the
same initial conditions after $7\times 10^5$ collisions for an
event-driven simulation with non-contact interactions, where all
details are given in the following sections. We use
relatively weak potentials where the ratio of the potential at contact
to the initial granular temperature is $-10^{-5},0,10^{-5}$, for
attractive, neutral, and repulsive potentials, respectively. The three systems
present similar clusters in shape but their evolutions and structures
are different; in particular, clusters for the attractive case are
denser than in the hard-sphere case, while for the repulsive case they
are relatively more dilute.
The following section introduces the theoretical mean field
approach we will use to study the homogeneous cooling
regime that occurs before the clustering.

\section{Dissipation rate modification due to attraction and repulsion}
\label{sec:coolingRate}
Haff's law is valid only for particles with hard-core interaction in the 
homogeneous cooling state (HCS). For
long range interaction, M\"uller and Luding \cite{muller11,muller09}
predicted, using a modified pseudo-Liouville operator formalism, a
reduced cooling rate due to the repulsive forces and an increased rate
due to attractive forces (extending the results of
\cite{scheffler2002}). In their theory, the ratio of the 
potential at contact, $\phi$ (see subsection\ \ref{sec:DiscPot}),
to the temperature is the control parameter,
\begin{eqnarray}
\Gamma =\frac{| \phi |}{T_g}~, 
\end{eqnarray}
with different sign convection as in Ref.\ \cite{muller11},
where $\Gamma^*=\phi/T_g$ was used in some equations. Since the theory is a 
simple mean field theory, the shape of the potential does not enter in the
formulation, only its value at contact \cite{muller11}.

Since the cooling rate is modified, the transport coefficients of the
system will be modified accordingly. We will focus only on the change
produced in the dissipation rate, since it is the dominant term controlling
the dynamics of the system. Namely, the modified dissipation rate is:
\begin{eqnarray}
     I = I_0 \psi(\Gamma)
       := I_0 \left\{
     \begin{array}{ll}
       1  &: {\rm Haff}\\
       \exp\left(-\Gamma \right) &: {\rm Repulsive}\\
       \left( 2 - \exp\left(- \Gamma \right) \right) &: {\rm Attractive}
     \end{array}
   \right.
\label{eq:Gamma}
\end{eqnarray}
These modifications were derived for 3D systems by computing the
average effect of a long-range potential in the collision frequency of
a granular system with a pseudo-Liouville operator approach. We will
use them directly in 2D accounting for the different dimensionality in
the pair correlation function at contact and the numerical factors in
the Enskog collision time. This is justified -- and confirmed by our
simulations -- since the integration of the Liouville operator
considers collisions in a plane due to angular momentum conservation,
and hence only the prefactor is different between 3D and 2D, while the
functional form in $\Gamma$ remains identical.

\begin{figure}[htp!]
  \epsfig{file=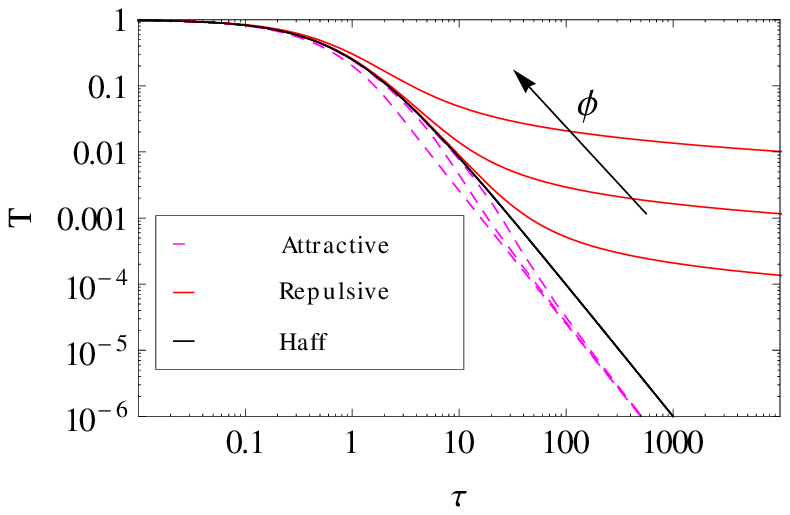,width=.475\columnwidth}
  \hfill
  \epsfig{file=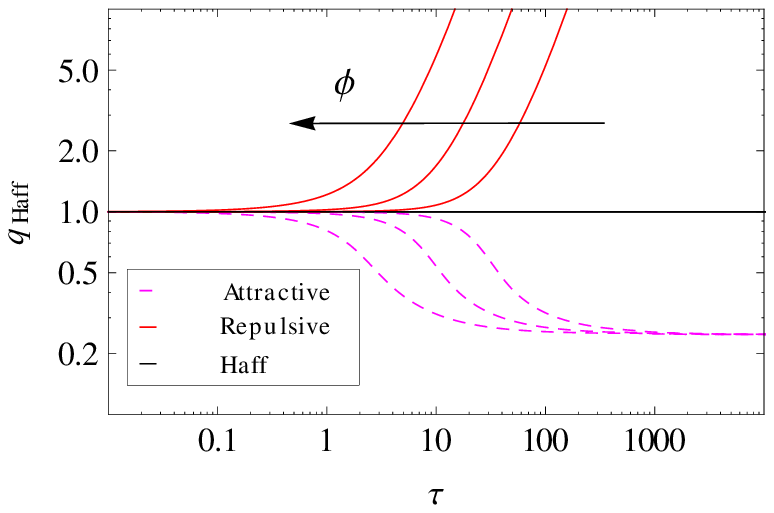,width=.475\columnwidth}
\caption{Cooling of granular systems. Left: temperature evolution for
  hard-spheres (Haff's law, solid black), repulsive interaction and
  attractive interaction as given by the numerical solution of the
  temperature equation \eqref{eq:PDEenergy2} considering the
  dissipation term as in Eq.\ \eqref{eq:Gamma} for different values of
  $\phi = 10^{-3},10^{-2},10^{-1}$. Right: quality factor $q_{\rm
    Haff} = T/T_{\rm Haff}$ for the same long range potentials as on
  the left plot.}
\label{fig:cooling}
\end{figure}

The homogeneous cooling with long range interaction is given by
Eq.~\eqref{eq:PDEenergy2} taking all the spatial derivatives equal to
zero and using the modified cooling rate from
eq.\ \ref{eq:Gamma}. This was numerically solved (we used Mathematica
8) and Fig.~\ref{fig:cooling} shows the evolution of the temperature
for three cases: attractive ($\phi<0$), repulsive ($\phi>0$) and no
long-range interaction (Haff's law, $\phi=0$).

Physically, Fig.\ \ref{fig:cooling} says the following: In Haff's
case, the dynamics of the systems becomes slower as time advances,
making the dissipation slower, and so on, as long as the system is
homogeneous \cite{haff83}. In the presence of non-contact forces, the
system can have at least two different regimes: at the beginning, the
thermal energy is larger than the repulsive/attractive energy, its
effect being negligible. As the system cools down, the
repulsive/attractive barrier will start to be felt, and the cooling
will be consequently modified. For the attractive force the prediction
says that once the particles start to feel the attraction they will
dissipate energy faster but nevertheless will retain the power law of
the dissipation and keep the homogeneity.

\section{Event-Driven simulations}
\label{sec:simulations}

Event-driven simulations have been widely used to study granular gases
\cite{mcnamara93b,luding99,miller04t,miller04b,luding05e,luding09,poschel05b}, 
and have shown to capture the correct behaviour when compared to kinetic theory. 
In what follows we present the details of the simulation algorithm for
discrete potential simulations where the range of attraction is finite.

The simulations used to prove the theoretical predictions in
Ref.\ \cite{muller11} were done with continuous potentials, which make
them computationally expensive and time consuming. However, the
theoretical description does not consider the shape of the potential
but only its repulsive barrier. Thus we complement the original
theoretical and numerical work by considering discrete potentials,
as readily simulated by event driven algorithms \cite{miller04t}
that are typically much faster but -- to the knowledge of the authors
-- could not yet be parallelised as efficiently as continuous
potential simulations \cite{miller04t}.

\subsection{Discontinuous potentials}
\label{sec:DiscPot}
Discrete potentials, such as the hard sphere model, have an important
advantage over more complex ``soft'' potentials. Between collisions
the spheres or molecules experience no forces and travel on ballistic
trajectories. The dynamics can be solved analytically, and the
integration of the equations of motion is processed as a sequence of
events rather than by fixed, small time-steps. Current event driven
molecular dynamics algorithms are quite advanced and allow the
simulation of large systems for the long times required to extract
accurate transport properties and study, e.g., the evolution of
clusters \cite{gonzalez2010,gonzalez2011} over many orders of magnitude.

\begin{figure}[htp!]
\centering
\epsfig{file=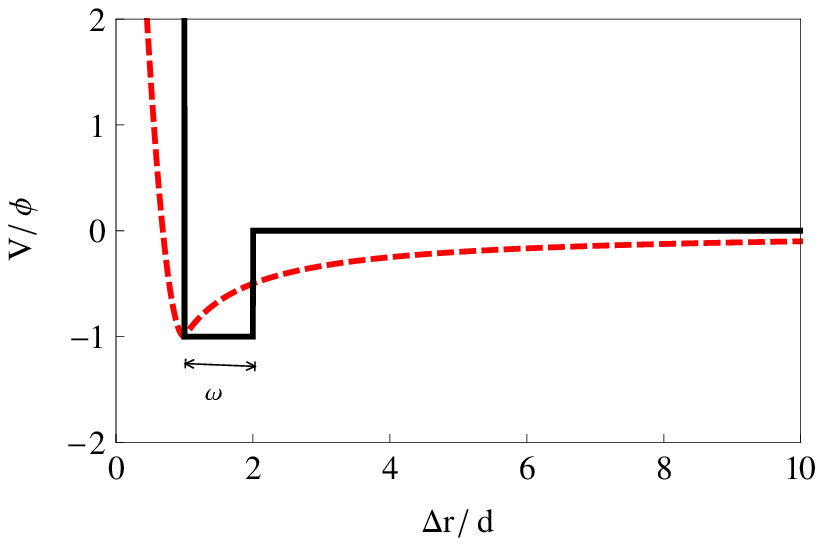,width=.475\columnwidth}
\hfill
\epsfig{file=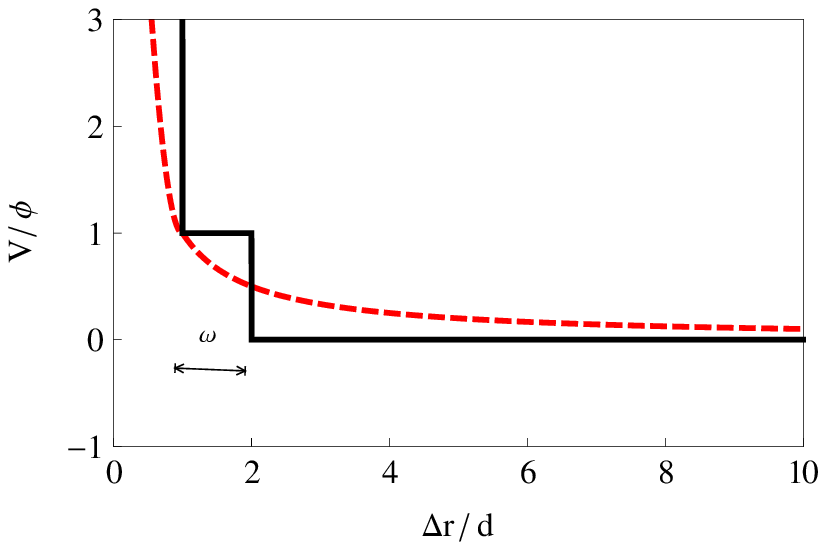,width=.475\columnwidth}
\caption{Plot of the potential energy for both continuous and
  discontinuous models with attractive (left) and repulsive (right)
  potentials as a function of the inter-particle distance $r$.}
\label{fig:EDpotential}
\end{figure}

For our simulations we use a two-level potential, see
Fig. \ref{fig:EDpotential}. The core is a hard sphere and there is a
well (or barrier) at a distance $\omega = 1.5$ (for all simulations
unless otherwise noted) and of amplitude $\phi$. 
At collisions of the hard sphere cores, the relative velocity is updated
only in the normal direction according to
\begin{eqnarray}
  \mathbf{v}'_{ij} = \mathbf{v}_{ij} 
                   - (1+r) \left ( {\mathbf{v}_{ij} \cdot \hat{\mathbf{n}}_{ij}}\right ) 
                                                          \hat{\mathbf{n}}_{ij} ~,
\end{eqnarray}
with normal unit vector $\hat{\mathbf{n}}_{ij}=\mathbf{r}_{ij}/|\mathbf{r}_{ij}|$,
where $\mathbf{r}_{ij}$ is the distance between the particles $i$ and $j$,
$\mathbf{v}_{ij}$ is their pre-collisional relative velocity, and the
prime indicates a post-collisional quantity. At crossing the well
(barrier or sink) the particles lose or gain energy instantaneously, 
depending on the direction of their relative motion, for details see
Ref.\ \cite{miller04t}.  The normal velocity after crossing is:
\begin{eqnarray}
  {v}^*_{n} = \sqrt{v^2_{n} \pm 2\phi/m}~,
\end{eqnarray}
with the normal velocity before, 
$v_{n}=\mathbf{v}_{ij} \cdot \hat{\mathbf{n}}_{ij}$, 
and where the sign depends on whether the
potential is attractive or repulsive, whether the collision is
outgoing or incoming, and if $v_n$ is large enough to cross the
well or whether the particles are reflected similar to the hard
core collision. For example, in the case of a rapidly incoming collision, 
in the attractive (repulsive) case, the particles gain (lose) energy, 
while for an outgoing collision this is reversed. How the sign and magnitude of this
potential affect the macroscopic evolution of a free-cooling gas is
the subject of the rest of the study.

\subsection{System and Preparation}
The simulation consists of a system of $N$ particles in a square in 2D
(or cube, in 3D) of side length $L$ with periodic boundary
conditions. Particles are mono-disperse with diameter $d$ and mass
$m$. The packing fraction of the system is given by $\nu = N\pi
d^3/(6L^3)$ in 3D and by $\nu = N\pi d^2/(4L^2)$ in 2D.  The
simulations were carried out in DynamO, a free and open-source
event-driven code \cite{bannerman11}.

\subsubsection{Initial Conditions}
The initial state is prepared as follows: (1) Start with a square
lattice (2D) or a hexagonal close packing (3D), uniform random velocities 
and a given $\phi$. (2) Let the system equilibrate, while each particle collides
at least 100 times elastically, until a homogeneous regime is
reached. This is tested by looking at the density and
the distribution of velocities and confirming that they are homogeneous 
and very close to Maxwellian, respectively. (3) Once thermalised,
the velocities are scaled so $T_0 = 1$ and dissipation is turned
on. From here, the system is allowed to freely cool down with a fixed 
particle coefficient of restitution, $r$.

\subsubsection{Simulation Units}
The simulations are performed using a non-dimensionalised system. 
The units of length, mass and time are set such that, 
$d=1$, $m=1$ and $T_0=1$. Since we are interested in the perturbative 
case, the potential, $\phi$, is mostly varied in the range from
$10^{-5}$ to $10^{-3}$ for both, attractive and repulsive cases.

\section{Numerical Results}\label{sec:results}
In this section we present first the results for the attractive
regime, followed by the repulsive case results.

\subsection{Attractive forces}
In this subsection we analyse data for cooling with attractive
potentials of different intensity and for different coefficients of
restitution. The objective is to understand how the dynamics is
influenced by these two factors and how they interact. We focus on the
evolution of three aspects: the temperature, i.e.\ the degree of cooling; 
the velocity distribution of the particles; and, the cluster structure and
size distribution.

The initial state for different systems is always the same: the
distribution of particles is homogeneous, with a Maxwellian velocity
distribution, and, hence, the cooling is well described by Haff's
law. As time passes dissipation and the attractive potential will have
time to act and modify this picture giving rise to a modified
cooling dynamics and clusters.

\subsubsection{Cooling}
The theoretical prediction from M\"uller and Luding \cite{muller11}
for the cooling rate is a homogeneous state with a twice as large 
dissipation rate. From Eq.\ \eqref{eq:Gamma}, we have for $t\to \infty$ 
the temperature $T_g \to 0$ and thus the control parameter $\Gamma \to \infty$ 
so that $I = 2 I_0$.  Furthermore, the theory predicts
that systems with different $|\phi|$ will deviate from Haff's law at
different times; the smaller $|\phi|$ the more time it takes to
deviate, however, the functional form remains the same
\footnote{By
functional form we mean that changing the potential is equivalent to make
a change of variable in time.}.
  
\begin{figure}[htp]
  \centering

  \epsfig{file=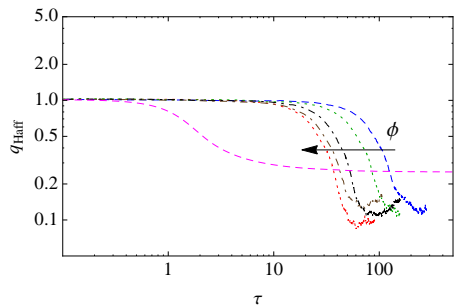,width=.475\columnwidth}
  \caption{Temperature evolution normalised by Haff's law with
    attractive potential for $r=0.99$, $\phi
    =10^{-3},2\times10^{-3},5\times10^{-3},8\times10^{-3},10^{-4}$
    together with Haff's law (solid-black) and the theoretical
    prediction for $\phi = 1$ (dashed magenta). The onset of cooling
    in the simulations is well predicted but the later cooling rate is
    overestimated, the magenta line runs noticeably on top of the
    simulations.}
  \label{fig:coolingAtt}
\end{figure}

Figure \ref{fig:coolingAtt} shows the temperature (normalised by Haff's
law) for systems with $\phi
=10^{-3},2\times10^{-3},5\times10^{-3},8\times10^{-3},10^{-4}$ and low
dissipation so the cooling is homogeneous. The cooling is well
predicted by the theory only for the initial deviation from Haff's law
but there is no agreement after that; simulations dissipate more
energy than predicted, since the theory cannot account for the
inhomogeneities in the simulations.  
Interestingly, the simulations do seem to
follow power laws of similar slope but with different scaling
factors. This can be seen around $\tau\simeq 100$ where all the curves
saturate close to $q_{\rm Haff} \simeq 0.1$. Due to the long time 
it takes to simulate the clusterised state, we do not have
data in the very long time regime.

Snapshots of the evolution for one of these low-dissipation systems
can be seen in Fig.\ \ref{fig:55}. Once the attractive force is larger
than the thermal fluctuations, the system develops clusters. The
structure of these clusters is typical for a cluster-cluster
aggregation process, with an exponential decay on the cluster size
(see below).

\subsubsection{Velocity Distribution}
The temperature alone is not a good indicator of the structure
formation process since it also depends on $\phi$. A better indicator
is to look for the deviation of the velocity distribution from a
Maxwellian. For this, we focus on the evolution of the kurtosis
$\beta_2= \mu_4/\mu_2^2$, where $\mu_i$ denotes the $i$th central
moment (and $\mu_2$ in particular the variance) as a function of
$\Gamma$. When the $\beta_2\simeq 3.108$ the system is
homogeneous. (It must be noted that since the system has a finite
number of particles, the kurtosis does not reach the theoretical value
but fluctuates near to it.) As soon as clusters appear in the system,
the velocities of the particles in the cluster are more correlated and
hence the distribution of velocities deviates from the Maxwellian
towards higher values of $\beta_2$.

By studying the evolution of the system as a function of the relevant 
control parameter, $\Gamma$ (i.e.\ the strength of the non-contact potential
relative to the granular temperature, see Eq.\ \ref{eq:Gamma}), one
is looking at the same time at the temporal evolution of the
system, since $T_g$ is decreasing with time. 
The advantage of looking at $\Gamma$ instead of just at the true
time of the system is that different coefficients of restitution can be
compared in one plot without having to scale the temporal axis,
see Fig.\ 2 and Eq.\ (4.2) in Ref.\ \cite{muller11}. 

\begin{figure}[htp]
  \centering
  \epsfig{file=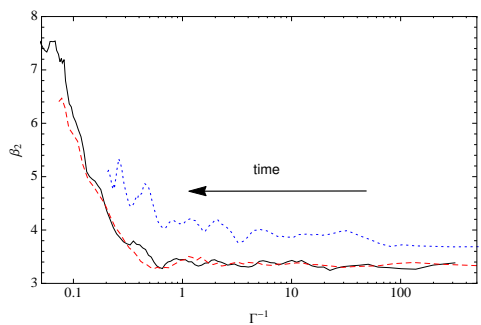,width=.475\columnwidth}
  \caption{ Kurtosis of the velocity distribution as a function of
    $\Gamma$ for systems with attractive potential. Different $\phi$
    collapse on the same curve as long as the coefficient of
    restitution is large enough.  When the coefficient of restitution
    goes below the critical value given by HCS, the system develops
    dissipative clusters before the attractive cooling sets in. The
    bottom lines correspond to an evolution as in Fig.\ \ref{fig:55},
    while the upper line corresponds to an evolution as observed in
    Fig.\ \ref{fig:56}.}
  \label{fig:gammaKurtosisAtt}
\end{figure}

Figure \ref{fig:gammaKurtosisAtt} shows the kurtosis during the
cooling for three systems with attractive potentials for different
coefficients of restitution and potential strength, at a fixed
density. If the cooling is homogeneous (solid and dashed lines), the
deviation from the homogeneous state is due solely to the attractive
potential and sets up when $\Gamma^{-1}$ is smaller than one. When the
cooling is not homogeneous, i.e.\ inelastic clusters appear, the
system deviates from the Maxwellian distribution of velocities before
the attractive potential has time to act. This indicates that there
are two mechanisms of clustering present in the system: inelastic
cluster formation and fractal-like aggregation.  Depending on $\Gamma$
one or the other will dominate.

\subsubsection{The cluster structure as a function of $\phi$}
Roughly, we divide the phenomenology in two different regimes
quantified by $\Gamma$: comparable ($\Gamma \sim 1$) and small
($\Gamma \ll 1$) short-range effects. In the first regime, the
attractive potential is stronger than the kinetic energy and hence
when particles collide they tend to remain bound together creating
clusters in a homogeneous way; this is the fractal-like process we
discussed before. The second regime is the one where the attractive
potential is weak enough to let the cluster instability appear in the
system; this is the inelastic cluster regime. However, once a cluster
occurs, the attractive force binds it together and makes the
dissipation rate to be stronger than in an equivalent cluster of hard
spheres. The regime $\Gamma \gg 1$ is not considered, since it
resembles the limit of cluster-cluster aggregation process
\cite{Meakin1984} and the dissipation has no role to play there.

\begin{figure}[htp]

  \epsfig{file=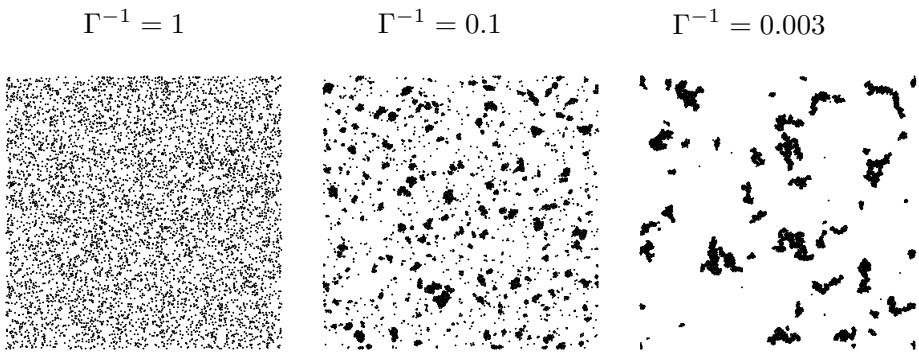,width=.475\columnwidth,angle=-90,clip=true,trim= 4cm 0cm 4cm 0cm} 
  \caption{ Simulation snapshots at different $\Gamma$ (time increases
    from left to right) for a considerable attractive potential ($\phi
    = -1$) for $N = 6400$, $\nu=0.0578$ and $r=0.99$. This corresponds
    to a system equivalent to the lower dashed line in
    Fig.\ \ref{fig:gammaKurtosisAtt} but with a larger attractive
    force.}
  \label{fig:55}
\end{figure}

\begin{figure}[htp]
\centering 
 \epsfig{file=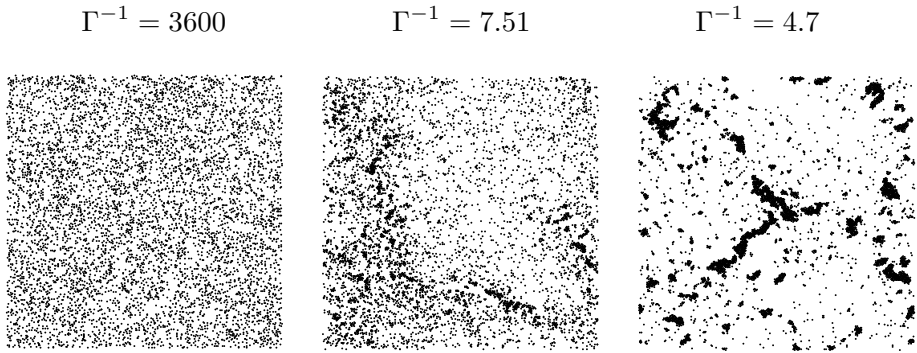,width=.475\columnwidth,angle=-90,clip=true,trim= 4cm 0cm 4cm 0cm} 
  
  \caption{Simulation snapshots at different $\Gamma$ (time increases
    from left to right) for a weak attractive potential ($\phi =
    -10^{-4}$) for $N = 6400$, $\nu=0.0578$ and $r=0.6$. This
    correspond to an evolution as the one depicted in
    Fig.\ \ref{fig:gammaKurtosisAtt}, upper dotted line.}
  \label{fig:56}
\end{figure}

Figures \ref{fig:55} and \ref{fig:56} show snapshots of systems for
decreasing $\Gamma$ from left to right, and a different coefficient 
of restitution and $\phi$ in each figure. In the first case,
Fig.\ \ref{fig:55}, with a relatively strong potential and low
dissipation, the cooling is homogeneous until $\Gamma \simeq 1$ when
clusters start to appear homogeneously, in a way similar to the
cluster-cluster aggregation process. Fig.\ \ref{fig:56} shows
snapshots for a system with an increased dissipation,
$r=0.6$ and reduced attractive potential, $\phi=-10^{-4}$. This gives
enough time for dissipation-induced clusters to appear. Once a cluster
appears, since the relative velocity of its particles is smaller than
the thermal fluctuations, the attractive force come into play, further
increasing the dissipation inside the cluster. The effect of this is that
there is no energy left to break the cluster, and once it forms, it
will remain in the system with roughly the same shape, thus creating a
non-homogeneous clusterisation. This can be seen from the cluster size
distribution and its temporal evolution as well as from the snapshots.

\begin{figure}[htp]
  \centering
  
\epsfig{file=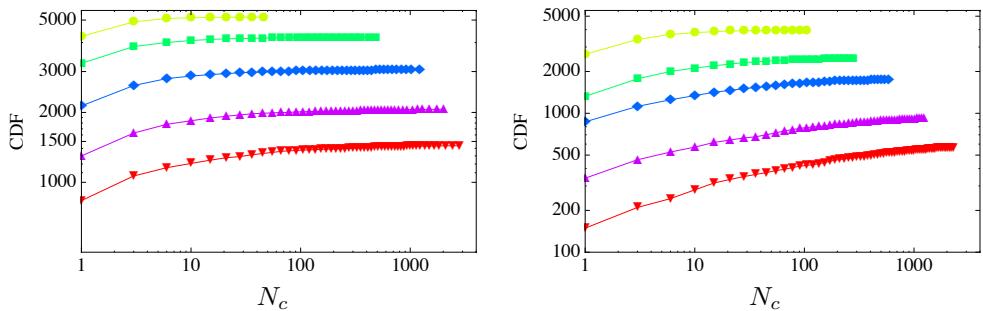,width=.425\columnwidth,angle=-90,trim=5.5cm 0 4.5cm 0, clip=true}  
\caption{Cumulative cluster size evolution as a function of cluster
  number $N_c$ (number of particles in a given cluster \cite{luding99}) 
  for a homogeneous clustering system (left) and a system with
  inelastic cluster formation (right). In both plots different symbols
  represent different stages in the evolution from more homogeneous
  (upper lines) to clusterised systems (bottom line).}
  \label{fig:CDFatt}
\end{figure}

\begin{figure}[htp]
  \centering
  \epsfig{file=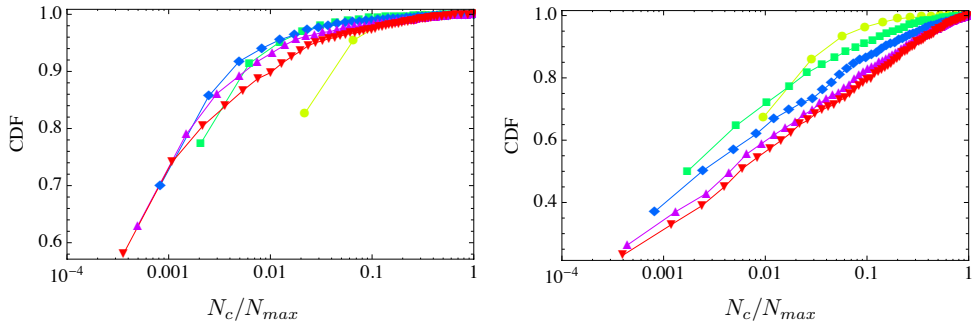,width=.425\columnwidth,angle=-90,trim=5.5cm 0 4.5cm 0, clip=true}   
  \caption{Cumulative cluster size evolution normalised by the largest
    cluster size, $N_{max}$. Same data as in Fig. \ref{fig:CDFatt}. For the
    homogeneous case the evolution of the size distribution seems to
    converge to a straight line.}
  \label{fig:CDFattNorm}
\end{figure}

Figure \ref{fig:CDFatt} shows the cumulative distribution function
(CDF) for the evolution of the cluster size distribution for two
systems, one with homogeneous clusterisation and the other with
inelastic cluster formation. In the homogeneous case (left) the
distribution of cluster sizes is broader than in the clusterised case
(right) and there are almost an order of magnitude fewer clusters of
size one for a similar maximum cluster size. On the other hand, for
the inelastic cluster formation (right), the distribution of sizes is
more heterogeneous: there are plenty of clusters of small size and a
few very large clusters. This can be seen also in the normalised plot,
Fig. \ref{fig:CDFattNorm}, where for the homogeneous case the size
distribution approaches a straight line, i.e.\ a logarithmic
function. This means that in the limit of infinite particles, there
are particles of all the sizes but the probability of finding a
cluster of size $n$ scales as $1/n$. In the case of inelastic
clusters, the evolution of the size distribution does not follow a
clear trend.

\subsubsection{Attractive cooling systems are always unstable}
Hard-spheres with attractive long-range interactions present
condensation \cite{vanKampen64}. Indeed, for the elastic case, there
is a region of temperatures where the liquid and gas phases can
coexist. Consequently, for a homogeneous gas of cooling hard-spheres
with long-range attraction, the temperature eventually reaches the
critical point where the liquid phase becomes stable, and then
clusters develop. These clusters will grow due to the inelasticity of
the particles to eventually reach a single cluster composed of all the
particles of the system. This is independent of the density and size
of the system. In this way, the free cooling of hard-spheres with
attractive wells is always unstable to cluster formation
disregarding the size of the system. However, the clusterisation
process can be, in turn, homogeneous or inhomogeneous. It must be
noted that the term homogeneous is scale-dependent: once the energy of
the system is low enough particles merge into pairs breaking the local
homogeneity, but remaining homogeneous in larger scale that comprises
several small clusters.

This makes a fundamental difference with the free cooling of
hard-spheres, where the clustering instability depends on the
wavelength of the perturbation; if the system is small enough, the
cooling will be always stable. We proved this by realising small
($N\approx 100$) simulations of attractive elastic particles -- the
phase separation always appeared independent of the system
size. Furthermore, even for two particles, if the temperature is low
enough, they will eventually find each other and merge into a cluster.
Thus, since the phase separation does not depend on the system size,
it always appears in a free cooling system when the temperature gets
low enough ($\Gamma \approx 1$).

\subsubsection{Comparison with wet granulates}
Our model can be compared to the cooling of wet granulates, see
Refs.\ \cite{Zaburdaev2006,Ulrich2009}, where the authors studied cooling
by using a very simple model for the interaction of two wet grains, 
which only accounts for the essential features of a capillary 
bridge: hysteresis and dissipation with a well-defined energy loss. 
Cooling 
is controlled by the probability for a bridge to break and hence
logarithmically slow in the long time limit, when a percolating
structure has been formed. In contrast to theirs, our model has the
dissipation occurring at collision and conserves energy at the
crossing of the energy barrier. 
These two microscopic
differences may have a great influence on the macroscopic behaviour of
the system. As we have seen, our model develops inelastic clusters,
contrary to Ulrich's \cite{Ulrich2009}, which is homogeneous. 
This could be due
to the strong attraction they use as initial condition ($\Gamma =
1$). A detailed comparison of the two models is beyond the scope of
this study; however, it remains as an interesting open question to
investigate in the future.

\subsubsection{Effects of higher density}
For the continuous potential of Ref.\ \cite{muller11}, the cooling of
denser systems was shown to be not predicted by the dilute limit
theory. This was due to the multi-particle interactions that occur in
dense systems with long-range coupling. To see if the discrete
potential reproduces this behaviour, we realised simulations analogous
to the ones in M\"uller and Luding's paper \cite{muller11}.

\begin{figure}
\centering
\epsfig{file=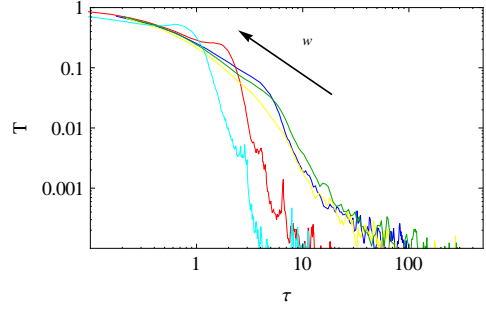,width=.5\columnwidth}
\caption{Cooling for dense ($\nu=0.157$) systems with large
  dissipation, $r=0.85$ and different potential width
  $w=1.1,1.5,2,2.5,3$ but fixed $\phi=0.1$. As the well becomes wider,
  the kinetic energy presents a peak due to large-scale
  reorganisation.}
\label{fig:denseCoolingW}
\end{figure}

One of the most notable features of the continuous potential
simulations is that the cooling is not monotonically decreasing: the
system phase-separates and the geometrical rearrangements produce a
temporary increase in the kinetic energy of the system. We did not
observe this increase in the temperature for the simulations with a
well width of $\omega = 1.5$ despite seeing the phase
separation. Thus we decided to vary the potential width to see if
this long-scale rearrangements are recovered with discontinuous
potentials. Figure \ref{fig:denseCoolingW} shows the energy for dense
systems with different well widths $\omega$ in 3D. For this density,
$\nu = 0.157$, all the systems present a phase transition below a
critical temperature: the homogeneous system becomes unstable and the
system phase-separates in a liquid and a gaseous (almost vacuum)
region. This is independent of the well width $\omega$.  However, as
the well becomes wider, the phase separation changes qualitatively,
from a percolating system with big bubbles to a system made of one big
drop. For different $\phi$ the qualitative evolution is similar, only
shifted to the corresponding temperature (data not shown). The
discrete system also presents a peak in the kinetic energy if $\omega$
is large enough. In this case, the qualitative change from a cooling
that is strictly monotonous to one that presents a peak, occurs around
$\omega_p \approx 2.25$. It must be noted that since the strength of
the potential was kept constant, the ``bump'' in the temperature
shifts to earlier times since the potential energy of the system
increases too when increasing $\omega$.

In systems where the density is so large that the particles'
interactions are not binary anymore, the theory can not predict the
cooling behaviour. However, some complex physics -- as the increase
in temperature due to geometrical rearrangements and structure 
formation \cite{muller08t} -- is recovered even with a discrete potential.

\subsection{Repulsive forces}
Intuition tells us that in the repulsive case there are two regimes
depending on whether the original hard-sphere system presents a
homogeneous or non-homogeneous cooling. For the homogeneous cooling
it is obvious that the repulsive potential will not enhance
the clustering and the cooling will remain homogeneous.  In the
non-homogeneous case, the repulsive forces will act against the
cluster formation since they tend to separate particles. However, one
can expect that if the potential is weak enough, it will not affect
the formation of clusters, at least temporarily. Eventually,
the temperature drops under the repulsive energy and the clusters are
eliminated. Quantifying this statement is the subject of the following
subsections.

\subsubsection{Homogeneous Cooling}

\begin{figure}[ht]
\epsfig{file=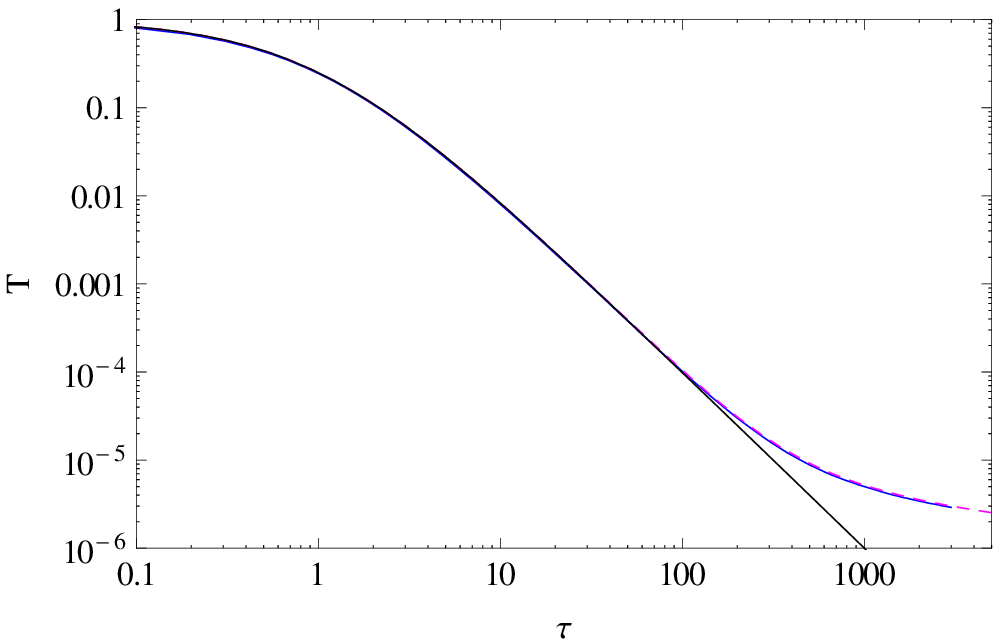,width=.45\columnwidth}
\hfill
\epsfig{file=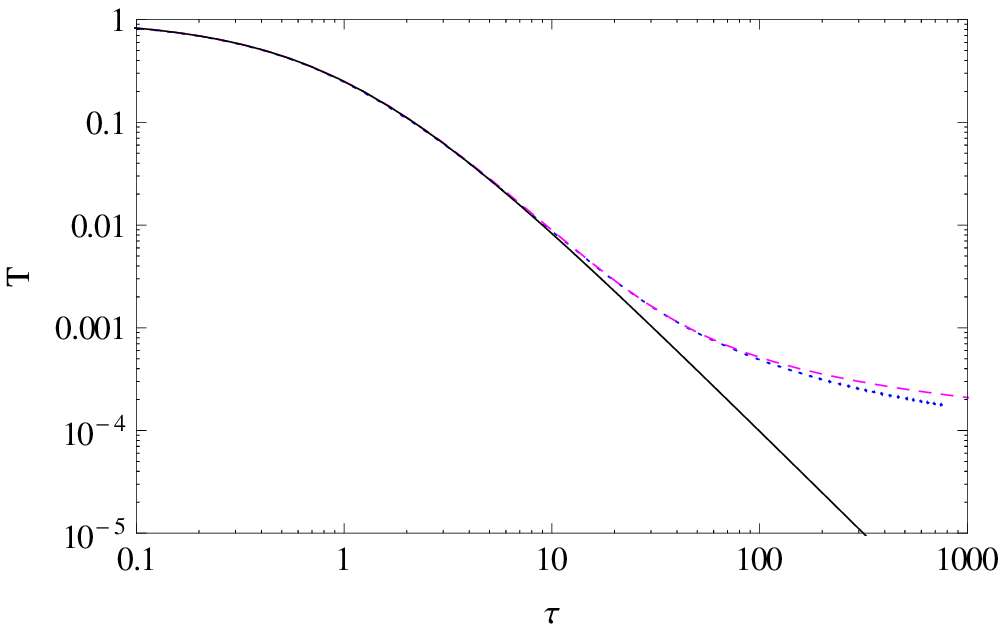,width=.45\columnwidth}
\caption{Cooling in 3D and 2D for the homogeneous case with $\phi
  =10^{-3}$ and $\phi =10^{-5}$, left and right respectively. The 2D
  system is denser ($\nu_{2D}=0.057$ versus $\nu_{3D}=0.0052$) and
  that may cause the small difference for large $\tau$ between the
  theory (magenta dashed line) and the simulations (blue dotted
  line). The black solid line is Haff's cooling state for reference.}
\label{fig:cooling2and3d}
\end{figure}

Contrary to the attractive case, the mean field theory for repulsion
is in great agreement with the simulations. The density remains
homogeneous and the temperature follows Eq.\ \eqref{eq:PDEenergy2}.
This is a strong result as there are no free parameters in the theory,
everything is determined by the potential at
contact. Fig.\ \ref{fig:cooling2and3d} shows the temperature for two
systems, in both 3D and 2D. The prediction works equally well for any
$\phi$, the only difference is that the cooling is shifted to
earlier/later times depending on the strength of the potential. For
3D, the modification in the cooling rate fits perfectly the cooling of
the system as long as it remains homogeneous. For 2D systems there is
an appreciable deviation for long times: the theory under-predicts the
cooling. The 2D system is denser ($\nu_{2D}=0.057$ versus
$\nu_{3D}=0.0052$) and that may cause the small difference for large
$\tau$ between the theory and the simulations.

\subsubsection{Non-Homogeneous Cooling}
A more interesting case is when the system is large/dissipative enough
to present clusters. If the repulsive potential is weak enough, the
time it takes to separate two particles that are close to each other
is much larger than the time the cluster formation takes to develop,
thus allowing for clusters in the system. With other words, if the potential 
at the beginning is much smaller than the thermal energy, the system has
time to develop clusters before the repulsive force separates
them. This is similar to the transient cluster formation observed
when there is a velocity-dependent coefficient of restitution
\cite{poschel05b}, and will be discussed in subsection \ref{sec:transient}. 

\begin{figure}[htp]
\centering
\epsfig{file=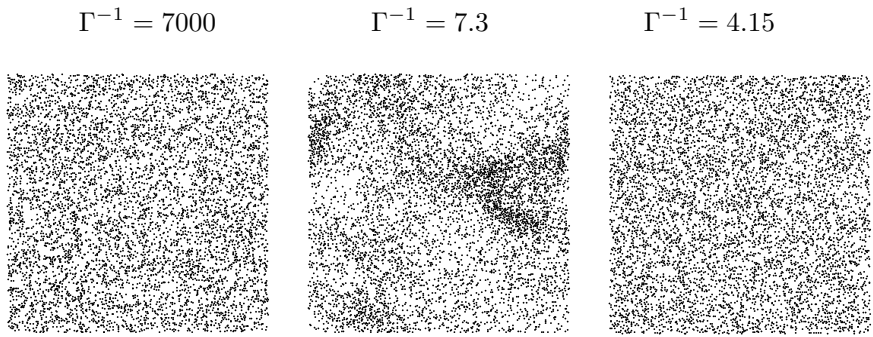,width=.45\columnwidth,angle=-90,clip=true,trim= 4cm 0cm 4cm 0cm} 
  
  \caption{Simulation snapshots at different $\Gamma$ (time increases
    from left to right) for a weak repulsive potential ($\phi =
    10^{-4}$) with $N = 6400$, $\nu = 0.057$ and $r=0.6$. In the
    middle picture the transient cluster formation can be seen;
    in this case, the kurtosis peaks to a high value due to
    the presence of clusters.}
  \label{fig:57}
\end{figure}

\begin{figure}[htp]
  \centering
  \epsfig{file=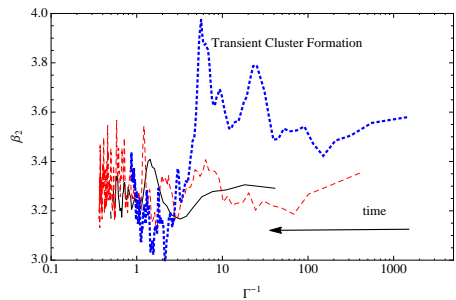,width=.45\columnwidth}
  \caption{Kurtosis of the velocity distribution for three systems
    with $\phi = 10^{-3},10^{-4},10^{-5}$ (solid, dashed and dotted
    lines) and different coefficients of restitution $r=0.99$, $0.95$, and $0.9$, 
    respectively, as functions of $\Gamma$. The peak in the kurtosis at 
    $\Gamma^{-1} \simeq 10$ corresponds to the appearance of a transient 
    cluster as the one from Fig.\ \ref{fig:57}, middle (where $r$ is different). 
    Before and after the peak correspond to left and right in the same figure. 
    As $\Gamma$ approaches unity, the velocity distribution tends to
    decorrelate and $\beta_2$ decreases.}
  \label{fig:gammaKurtosisRep}
\end{figure}

Figure \ref{fig:57} shows snapshots of the system at decreasing
$\Gamma$ from left to right, for a given repulsive potential and
coefficient of restitution.  In the middle picture, the transient cluster
structures can be seen, while before and after (left and right) the
system is homogeneous. This corresponds to the evolution of the system
represented in Fig.\ \ref{fig:gammaKurtosisRep} by the dotted blue
line.

Figure \ref{fig:gammaKurtosisRep} shows the kurtosis of the velocity
distribution as a function of $\Gamma$ for three different $\phi$ and
coefficient of restitution $r=0.99,0.95,0.9$. If the repulsive potential
is weak enough and there is enough dissipation, the cluster
instability can develop, making the distribution of velocities much
more correlated and hence having a larger kurtosis (dotted line). When
the thermal energy is comparable to $\phi$, i.e.\ $\Gamma \simeq 0.1$,
the repulsive force starts to destroy the clusters and produces again
a more Maxwellian distribution of the velocities. If the dissipation
is small enough, the evolution for different coefficients of
restitution in the plane $\Gamma-\beta_2$ is similar. The fluctuations
around $\Gamma =1 $ are due to the long time it takes the system to
reach lower temperatures, and hence the evolution appears compressed
when plotted as a function of $\Gamma$ instead of $\tau$. The
fluctuations look larger than in the repulsive case only because of
the different scales of plotting; for large $\Gamma^{-1}$ both
oscillate by a similar magnitude.

\subsubsection{Transient cluster formations}
\label{sec:transient}

The evolution of the free cooling system with repulsive interactions is
somewhat analogous to a system with a velocity dependent coefficient
of restitution, where clustering and shearing appear transiently
\cite{poschel05b}. However, the mechanism that controls the presence
of clusters is the repulsive force being non-zero and the relative
importance of dissipation and repulsion. Any repulsive force between 
the particles will eventually inhibit cluster formation (whereas
attraction and dissipation both work in favour of it). 
Since there is no external pressure to compress the system,
the particles always tend to separate from each other, where 
stronger dissipation slows down separation. However, this repulsion-related
mechanism has a temporal scale controlled by the intensity of the potential
at contact, and thus by the available ``escape kinetic energy'' 
$T_e \propto \phi/\Gamma$. 
For two spheres touching, the time they need to separate a
particle diameter, i.e.\ to not (mechanically) feel themselves anymore, 
is
$t_r \simeq d/\sqrt{2\phi/m}$. Equivalently, there is 
a shorter time-scale (for short-range potentials) that 
involves the range of the potential 
$t_{\omega} \simeq \omega d/\sqrt{2\phi/m}$\,
where $\omega$ is the width of the well in units of particle diameters,
which the particles need to leave the range of the potential.
Consequently, unless the time of free flight for the particles becomes 
comparable to or larger than $t_{\omega}$, the repulsive
potential is not able to separate the particles and the dynamics is
dominated by the hard-sphere properties. Once the collision 
time-scale of the system exceeds $t_{\omega}$ the clusters begin to 
disintegrate and the system eventually returns to a homogeneous state.

\subsubsection{Effects of higher density }

For dilute enough systems the cooling rate becomes practically
logarithmic, as is the case for a velocity-dependent coefficient of
restitution. However, if the density is high enough, the system can
reach an elastic regime due to the finite size of the system and the
consequent bound for the kinetic energy. For such systems, the
evolution is no longer independent of the coefficient of restitution,
as it is the case in the dilute regime (data not shown). Since the
barrier in the discontinuous case is abrupt, when the temperature is
low enough particles cannot overcome the repulsive barrier and stop
colliding. Then particles acquire an effective radius equal to the
well width and the energy of the system remains constant. 

Contrary to the attractive case, increasing the well width would be of
no use in trying to recover the physics of the continuous
potentials. What is needed in this case is to study how the inclusion
of more steps in the potential approaches the continuous
case. However, that is beyond the scope of this study.



\section{Phase diagram for cooling with non-contact interactions}
\label{sec:phaseDiagram}

The phenomenology for the cooling and clustering of hard spheres with
short- and long-range non-contact interactions of strength $\phi$
can be summarised as follows: 
for attractive potentials the freely cooling systems always
show clusters (or condensation) on the long term. If $\phi$ is 
small, relative to the granular temperature, the cooling is initially
homogeneous: however, due to the decrease of the granular temperature
(cooling), the relative importance of the non-contact interaction
increases and eventually leads to clustering. On the other hand, if 
$\phi$ is large enough, particle pairs stick to each other at 
first contact already and once pairs are bound together,
the cooling continues as a cluster-cluster aggregation process
\cite{gonzalez2010}. 
When a
system features a shear- or cluster-instability, it will
e.g.\ show inelastic clusters that cool down slower than the homogeneous
hard-sphere case \cite{luding99,miller04t,luding05e}, 
giving rise to a different dynamics and cluster structure. 
This slower cooling is not (and cannot be) predicted by the homogeneous 
mean-field theory that only takes into account the very weak interaction
limit case for very low densities.

\begin{figure}[ht]
\epsfig{file=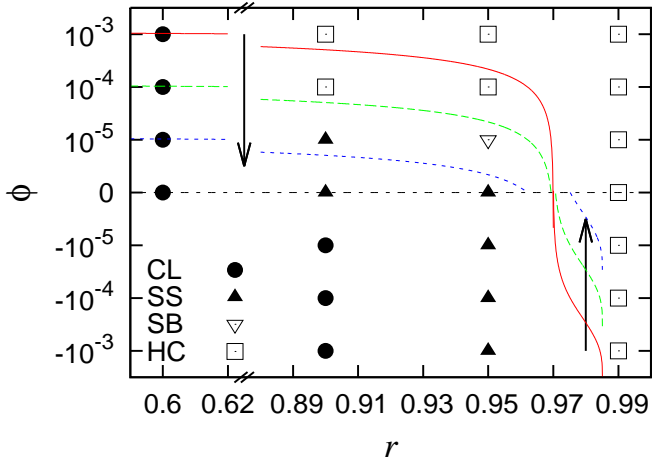,width=.5\columnwidth,angle=-90}
\caption{Phase diagram of free cooling 2D-systems for the potential strengths, $\phi$,
  where clustering (CL), strong shear (SS), shear banding (SB), and homogeneous cooling (HC)
  are reported, for different coefficients of restitution, $r$, with $N = 6400$ 
  and $\nu = 0.057$. For the corresponding hard-sphere systems (points), 
  the classical critical coefficient of restitution is $r_c = 0.97$, where
  the upper and lower lines meet. The lines are the transition curves for
  $T_g=10^{-3}$ (red), $10^{-4}$ (green), and $10^{-5}$, (blue).
  Since the scale is logarithmic, 
  we artificially cut-off at $|\phi|=10^{-6}$ and set the horizontal axis at 
  $\phi = 0$; hence not all the curves reach the critical value and thus do 
  not match exactly when crossing from positive to negative $\phi$.
  The arrows indicate the direction in which the transition lines sweep 
  through the system (towards smaller $|\phi|$).}
\label{fig:phase}
\end{figure}

For the repulsive case, the homogeneous cooling is well predicted by a
simple modification of the collision frequency, which implies a
modified dissipation rate. If the system is unstable to clustering and
the initial $\phi$ is small enough, the system will present inelastic
transient clusters, in a way analogous to a granular gas with a
velocity-dependent coefficient of restitution. Figure \ref{fig:phase}
shows a preliminary, qualitative sketch of the phase diagram, 
for both attractive and repulsive potentials at given density
and system size, for
different coefficients of restitution and potential strengths.

Qualitative (time-dependent) transition lines can be drawn on the phase 
diagram by computing the critical coefficient of restitution for the 
shear instability as a function of $\phi$, for a given system size
and density \cite{miller04t}. 
This approach is in analogy to the case of a velocity-dependent
coefficient of restitution from Ref.\ \cite{poschel05b}, where it 
was shown that classical stability analysis for the free-cooling gas
\cite{mcnamara93b} also holds for a time-dependent dissipation
rate. In particular, the eigen-value equation for the shear mode,
see Eq.\ (9) in \cite{poschel05b}, becomes time-dependent via the
dissipation rate. Stability analysis for the critical coefficient of
restitution of the shear mode in the dilute limit provides \cite{miller04t}:
$1-r^2 = (1-r_c^2)/\psi(\Gamma^*) = [\alpha/\psi(\Gamma^*)] k_{min}^2$\,,
where $r_c$ and $\alpha$ represent the classical result, involving
a combination of shear-viscosity and dissipation rate pre-factors,
valid when $\psi(\Gamma^*=0)=1$, with $\psi$ defined in Eq.\ (\ref{eq:Gamma}),
and $k_{min}$ being the smallest wave number possible
(the largest wave-length $L_{max} = 2\pi/k_{min}$ is the system size).
Due to the additional
non-contact force, the critical coefficient of restitution will
change, since it depends on time via $\psi(\Gamma^*)$, where we
have neglected a possible time-dependence of $\alpha$. 
%
The transition line in the plane $r-\Gamma^*$ is then:
\begin{eqnarray}
\Gamma^*_c = \Gamma_c = \log\left(\frac{1-r^2}{1-r_c^2}\right)~, 
\label{eq:Gamma_cr}
\end{eqnarray}
in the repulsive case, for $r<r_c$, while it is: 
\begin{eqnarray}
\Gamma^*_c = -\Gamma_c = -\log\left(\frac{1 - r^2}{1 - 2 r^2 + r_c^2}\right)~,
\label{eq:Gamma_ca}
\end{eqnarray}
in the attractive case, for $r>r_c$.
Here, $r_c$ is the classical result for the critical coefficient of restitution
in the given system without non-contact forces.
%

The {\bf phase space} can now be represented in two ways: (1) in the 
$r-\phi$ plane, where our simulations are fixed points,
but the transition line sweeps downwards with time, i.e.\ during
the evolution (cooling) of the system; or (2) in the $r-\Gamma^*$ plane,
where the transition lines are invariant, but the systems drift from
small to large $\Gamma=|\Gamma^*|$ during cooling.

(1)
Figure \ref{fig:phase} shows a phase diagram for 
different systems with fixed $r$ and $\phi$, together with the predicted 
transition lines $\phi_c = \Gamma^*_c T_g(t)$, for different granular
temperatures $T_g$, sweeping from large to small $|\phi|$ during cooling;
the (stable) homogeneous cooling state is predicted to the top and right of
the transition lines, as qualitatively well reproduced by the 
simulations.

%

In the repulsive case, a given system can be initially in the
stable (top-right) or unstable (bottom-left) zone. Since the
transition lines sweep (downwards) towards smaller $\phi$, the former 
systems will remain stable, whereas the latter will become stable eventually.
In the attractive case 
the unstable domain is larger; 
a system originally in the stable zone (right, close to $r_c$)
will eventually become unstable, since the transition line
sweeps (upwards) towards smaller $|\phi|$,
for $\phi<0$.

{\em
In summary, this renders repulsive freely cooling systems stable on the
long term, whereas attractive systems are expected to become unstable
eventually.}

\begin{figure}[ht]
\epsfig{file=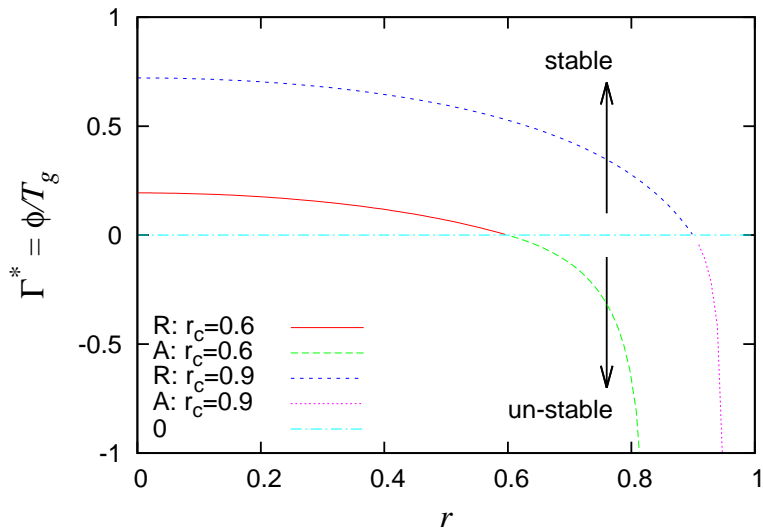,height=.8\columnwidth,angle=-90}
\caption{Phase diagram of free cooling systems for the control parameter
  $\Gamma^*$ and coefficient of restitution, $r$, with $r_c=0.9$ and $0.6$, 
  from Eqs.\ (\ref{eq:Gamma_cr}) and (\ref{eq:Gamma_ca}).}
\label{fig:phase2}
\end{figure}

(2)
Figure \ref{fig:phase2} shows the $r-\Gamma^*$ phase-space where
the transition lines are invariant and a freely cooling system
moves across them, as indicated by the arrows.  A repulsive system,
initially in the unstable zone will drift upwards and become eventually
stable, whereas an attractive system that could be initially stable
will drift downwards and eventually get unstable.

Other, more realistic systems, with a time- or velocity-dependent
coefficient of restitution \cite{mcnamara98,goldshtein03,poschel05b}
will also involve an additional side-wards drift of their state (data
not shown), whereas driven systems in a steady state would be truly
stationary in phase space (data not shown).

As final note, each system will require a certain time to react
to the hydrodynamic instability of its present state in phase space.
When systems drift, there can be a considerable delay time, e.g.\
when established clusters are expected to be destroyed in the stable
region.  More realistic systems might display transition lines that
are not just functions of $r$ anymore but could allow multiple
crossing points; however, this goes beyond the scope of this study.

\section{Conclusions and Outlook}\label{sec:conclusions}

The free cooling of granular matter with dissipation and -- in 
addition -- non-contact attractive or repulsive interactions has
been studied by means of event-driven simulations and mean-field
hydrodynamic theory. The systems homogeneous cooling
behaviour was compared to the modified mean-field theory
\cite{muller11}, where the corrected dissipation rate 
involves the ratio of the interaction potential strength and 
the granular temperature as the only new control parameter. 
This theory was developed for 3D; however, we have shown how
to apply it for 2D systems.
Simulations
with discrete potentials confirmed the theoretical predictions for low
densities in the repulsive case (in 2D and 3D) but fell short for the
attractive case where non-linear effects are more important.

In the present study, in contrast to previous studies \cite{muller08t,muller11},
we used different short-ranged well- or step-potentials \cite{miller04t,bannerman11} 
to complement the simulations with continuous long-range 
potentials.
Remarkably, the theory results that are used here do not have a 
potential range- or shape-parameter, only the potential magnitude is
needed as parameter -- at least for small $\Gamma$; 
since the theory was already tested with the longest range 
potential ($1/r$) one can imagine \cite{muller11}, we
consider the present results valid for both short- and long-range 
potentials.
This independence on shape and range was astonishing to the authors 
and suggests future research towards better theories that 
do consider shape and/or range of the interactions.

For {\em attractive potentials} the formation of structures, i.e., clusters,
is strongly enhanced, as is expected from the results on condensation
of elastic hard-spheres with attractive potentials
\cite{vanKampen64,muller08t}; this effect is not predicted by the simple
homogeneous mean-field theory. The geometry of the resulting clusters is
determined by a complex interplay between dissipation, density,
potential strength and range, and the granular temperature 
intensity that changes during time due to the ongoing cooling.

For {\em repulsive potentials}, the theory predicts consistently the cooling
behaviour for low dissipation and density as long as the system is
homogeneous. This confirms once again that discrete potentials are a
good approximation to smooth, continuous interaction potentials in
this regime and capture much of the interesting physics of particulate
systems. For larger dissipation and weak repulsion we have found that
transient clusters appear in the system in a way analogous to the
cooling of grains with a velocity dependent coefficient of
restitution \cite{mcnamara98,goldshtein03,poschel05b}. 
This is expected since the dissipation rate in the
hydrodynamic equations have a similar time dependence for the
free cooling in either case.

Finally, all results point to the great importance of some microscopic
parameters in the macroscopic evolution of granular systems and the
complex interplay of different micro-mechanic mechanisms that
lead to macroscopic, hydrodynamic phenomena. In particular, we showed
that the long-time dynamics for a system with both dissipation and
non-contact interactions is not simply the superposition of both
effects but depends in a more complex way on the particular evolution of 
the system, its initial state, and the particular parameters that
describe it.\\

\noindent
{\bf Future work:} 

The present study started from dissipative systems and added non-contact
interactions as new ingredient to predict a phase-diagram that encompasses
both mechanisms and their interplay (in the weak interaction limit). 
A very interesting situation,
not considered here, is self-gravity that can lead to clusters without 
any dissipation, so that dissipation represents a ``perturbation'', 
which surely affects the system evolution, as relevant e.g.\ in astro-physics,
and will complement the phase space in the limit $r \to 1$.

Future work on earth should consider also mixtures of different species 
of particles with different non-contact interactions as well as more 
realistic potential-shapes \cite{dammer04}; 
for example, magnetic (di-polar) interactions \cite{blair03,blair04}. 
For such potentials and for stronger interactions, relative to the fluctuations, 
it will be necessary to extended the simple theory used here. 
The present results are the basis for future research towards a more 
complete theory for long-range interactions in aerosols or suspensions.
%
Eventually, also driven
systems should be considered, see Refs.\ \cite{blair03,luding03c}
and references therein, since they can feature a steady-state
in time and should lead to a stable phase-diagram instead of the
time-dependent one observed above for cooling granular media.

Eventually, the understanding of the system evolution in its phase 
space combined with an experimental control of the interaction strength
and dissipation (e.g.\ by changing the surface- or solvent-properties)
will allow to better control and optimise engineering processes
by controlling at will either the transition lines or the system
position in phase space.\\

\noindent
{\bf{Acknowledgments}}\\
We thank H. J. Herrmann for many valuable discussions on dissipative
granular systems in the last decades, D. Wolf for additional discussions 
on systems with non-contact forces, and M. Bannerman 
for his support concerning the use of DynamO.
This study was supported by the Stichting voor Fundamenteel Onderzoek
der Materie (FOM), financially supported by the Nederlandse
Organisatie voor Wetenschappelijk Onderzoek (NWO), through the FOM
project 07PGM27, as well by the NWO-STW VICI grant 10828.

\bibliographystyle{unsrt} 

\bibliography{new,granulates3,biblio3}

\end{document}